\documentclass[a4paper]{spie_margins}

\usepackage{amsmath,amsfonts,amssymb}
\usepackage{graphicx}
\usepackage[colorlinks=true, allcolors=blue]{hyperref}

\usepackage{textcomp}    
\usepackage{xcolor}      
\usepackage{aas_macros}  
\usepackage{fix-cm}      

\graphicspath{{./images/}}  

\title{The GOTO Telescope Control System}

\author[a,b]{\mbox{Martin J. Dyer}}
\author[a,c]{\mbox{Vik S. Dhillon}}
\author[a]{\mbox{Stuart Littlefair}}
\author[d]{\mbox{Kendall Ackley}}
\author[e]{\mbox{Sergey Belkin}}
\author[f]{\mbox{Lisa Kelsey}}
\author[d]{\mbox{Tom Killestein}}
\author[g,d]{\mbox{Amit Kumar}}
\author[d]{\mbox{Joseph Lyman}}
\author[h,d]{\mbox{David O'Neill}}
\author[d]{\mbox{Krzysztof Ulaczyk}}
\author[d]{\mbox{Ben Godson}}
\author[a]{\mbox{Dan Jarvis}}
\author[d]{\mbox{Danny Steeghs}}
\author[e,i]{\mbox{Duncan K. Galloway}}
\author[j]{\mbox{Paul O'Brien}}
\author[k]{\mbox{Gavin Ramsay}}
\author[l]{\mbox{Kanthanakorn Noysena}}
\author[m]{\mbox{Rubina Kotak}}
\author[n]{\mbox{Rene Breton}}
\author[o]{\mbox{Laura Nuttall}}
\author[h]{\mbox{Ben Gompertz}}
\author[c,p]{\mbox{Jorge Casares Velázquez}}
\author[d]{\mbox{Don Pollacco}}

\author[ ]{\mbox{the GOTO Collaboration}}

\affil[a]{Astrophysics Research Cluster, School of Mathematical and Physical Sciences, University of Sheffield, Sheffield, S3 7RH, UK}
\affil[b]{Research Software Engineering, University of Sheffield, Sheffield, S1 4DP, UK}
\affil[c]{Instituto de Astrofísica de Canarias, E-38205 La Laguna, Tenerife, Spain}
\affil[d]{Department of Physics, University of Warwick, Coventry CV4 7AL, UK}
\affil[e]{School of Physics \& Astronomy, Monash University, Clayton VIC 3800, Australia}
\affil[f]{Institute of Astronomy and Kavli Institute for Cosmology, University of Cambridge, Madingley Road, Cambridge CB3 0HA, UK}
\affil[g]{Centre for Electronic Imaging, The Open University, Walton Hall, Milton Keynes MK7 6AA, UK}
\affil[h]{School of Physics and Astronomy, University of Birmingham, Birmingham B15 2TT, UK}
\affil[i]{Institute for Globally Distributed Open Research and Education (IGDORE)}
\affil[j]{School of Physics \& Astronomy, University of Leicester, University Road, Leicester LE1 7RH, UK}
\affil[k]{Armagh Observatory \& Planetarium, College Hill, Armagh, BT61 9DG, UK}
\affil[l]{National Astronomical Research Institute of Thailand, 260 Moo 4, T. Donkaew, A. Maerim, Chiangmai, 50180, Thailand}
\affil[m]{Department of Physics \& Astronomy, University of Turku, Vesilinnantie 5, Turku, FI-20014, Finland}
\affil[n]{Jodrell Bank Centre for Astrophysics, Department of Physics and Astronomy, The University of Manchester, Manchester M13 9PL, UK}
\affil[o]{Institute of Cosmology \& Gravitation, University of Portsmouth, Portsmouth PO1 3FX, UK}
\affil[p]{Departamento de Astrofísica, Universidad de La Laguna, E-38206 La Laguna, Tenerife, Spain}

\authorinfo{Send correspondence to MJD - email: martin.dyer@sheffield.ac.uk}

\begin{document} 
\maketitle

\begin{abstract}
The Gravitational-wave Optical Transient Observer (GOTO) operates a network of robotic, wide-field survey telescopes searching for optical transients to gravitational-wave signals and other astronomical events. Since 2023, the network has comprised two antipodal sites in the Canary Islands and Australia, each hosting two independent robotic mounts. The GOTO Telescope Control System (G-TeCS) is a custom Python software package that manages all operational aspects of the network; from alert processing and target scheduling to on-site hardware operations and monitoring. The central system consists of an alert monitor, observation database and scheduler, which issues targets to the four nodes. Each telescope operates using a system of independent control daemons, with a pilot master control program to coordinate nightly operations. This architecture, in development since 2015, provides a flexible and robust system that now successfully operates the fully autonomous network.
\end{abstract}


\newpage

\section{The GOTO Project}
\label{sec:project}

\begin{figure}[t]
    \begin{center}
        \includegraphics[width=0.48\linewidth]{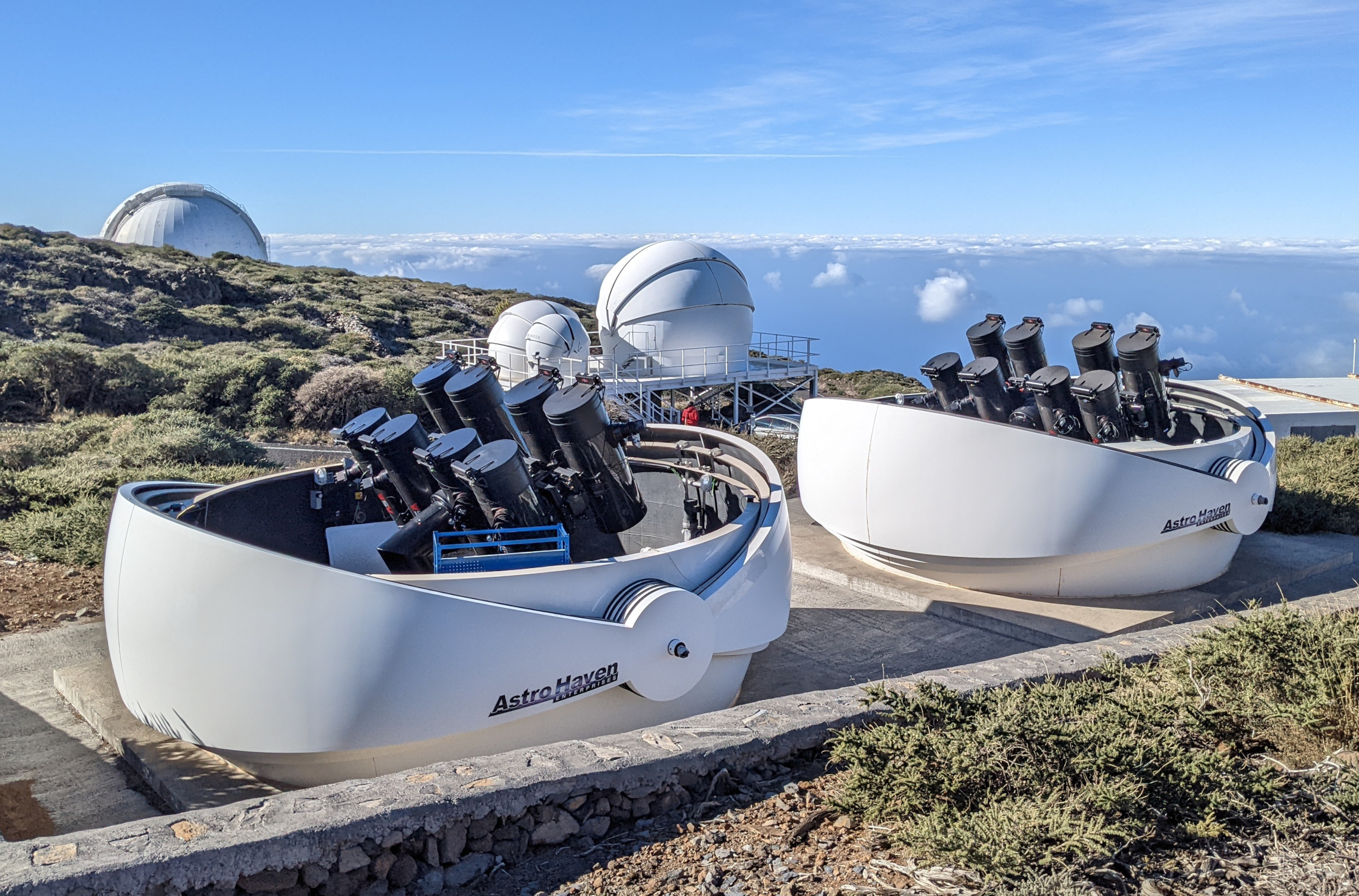}
        \includegraphics[width=0.48\linewidth]{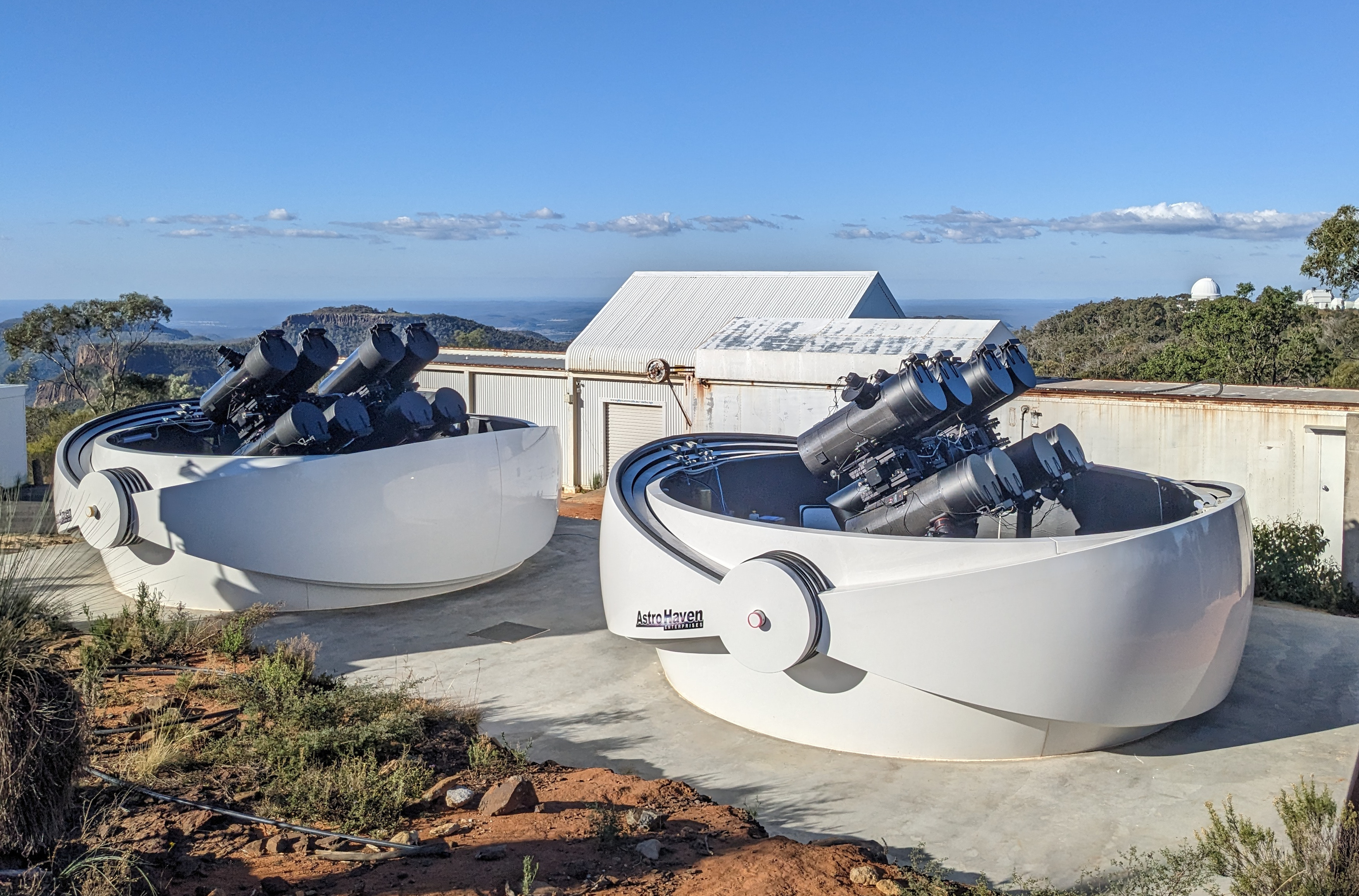}
    \end{center}
    \caption{
        The complete GOTO network as of April 2023. Left: GOTO-North on La Palma, with GOTO-1 on the left and GOTO-2 on the right. Right: GOTO-South at Siding Spring, with GOTO-3 on the left and GOTO-4 on the right.
    }\label{fig:photos}
\end{figure}

In recent years, a focus of astrophysics research has been on multi-messenger observations of astronomical events. While the electromagnetic spectrum (optical light, x-rays, $\gamma$-rays, radio etc) can tell us a lot about an event, the addition of other parallel "messengers" such as neutrinos, cosmic rays and gravitational waves allows different features to be explored and open new windows onto our understanding of the universe.

The \textit{Gravitational-wave Optical Transient Observer} (GOTO) is a network of robotic survey telescopes that are dedicated to responding to live alerts, particularly from gravitational-wave detectors, and hunting for any optical counterpart sources \cite{prototype}.

The GOTO network consists of four robotic telescopes shown in Fig.~\ref{fig:photos}: two on La Palma in the Spanish Canary Islands (together \textit{GOTO-North}) commissioned in 2021, and two at Siding Spring Observatory in Australia (\textit{GOTO-South}) commissioned in 2023. Each mount holds eight individual telescope tubes, which combine to give an overall field of view of 44 square degrees. When operating in survey mode the network can cover the entire visible sky every 2-3 nights \cite{goto2024}.

\begin{figure}[t]
    \begin{center}
        \includegraphics[width=0.95\linewidth]{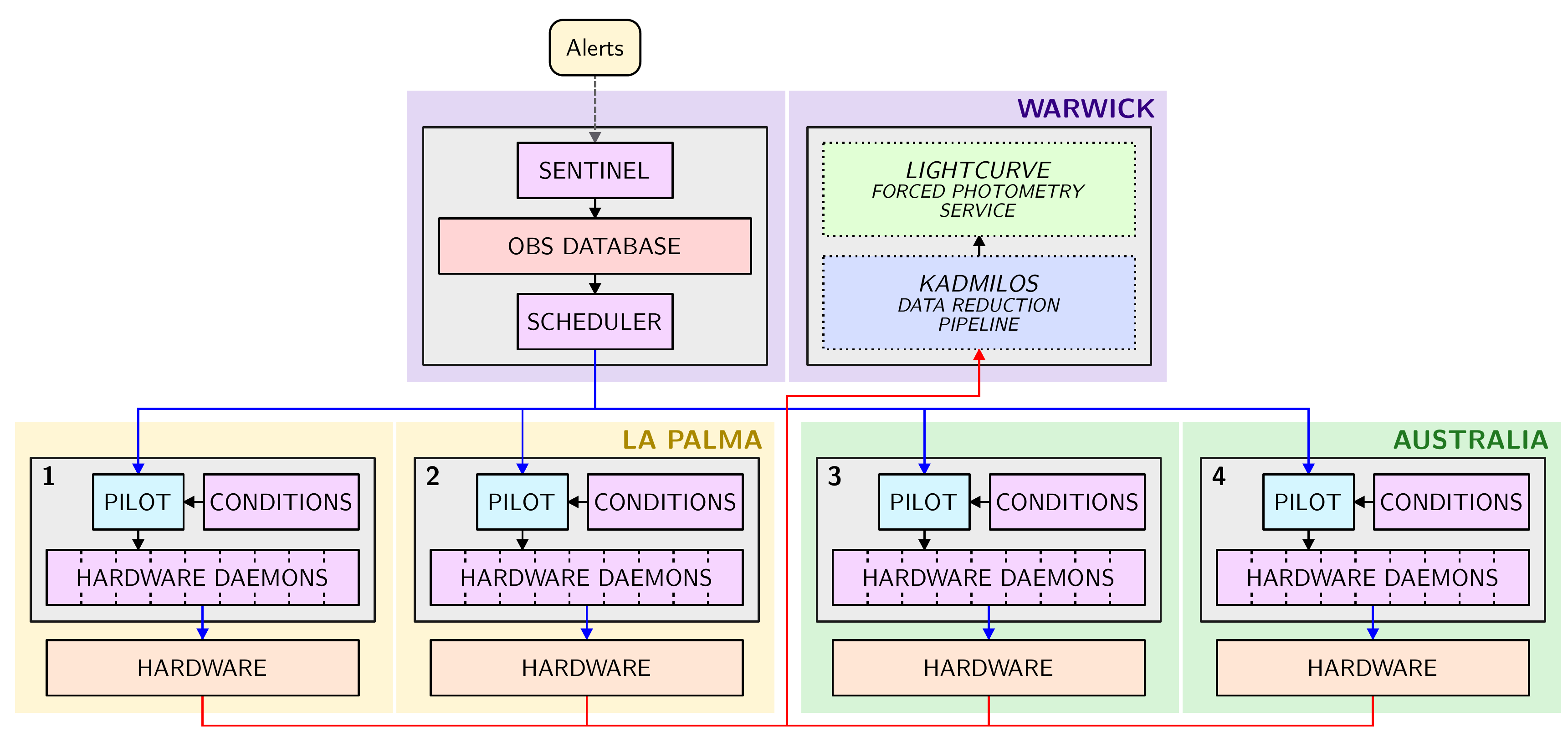}
    \end{center}
    \caption{
        A flow chart visualising the GOTO control network. Gravitational wave and other external transient alerts are received by the sentinel and inserted into the observation database, which the central scheduler uses to direct observations to the pilots of the individual telescopes. The output raw images are then transferred back to Warwick for processing by the Kadmilos pipeline.
    }\label{fig:flow}
\end{figure}

All four telescopes are operated as a single autonomous system by the \textit{GOTO Telescope Control System} (G-TeCS), depicted in Fig.~\ref{fig:flow}, which handles all aspects of the telescope scheduling and operations \cite{thesis,gtecs2020}.

\section{CENTRAL SCHEDULING}
\label{sec:scheduling}

The core of the G-TeCS control system is the central observation scheduling system from the \texttt{gtecs.obs} package \footnote{\url{www.github.com/GOTO-OBS/gtecs-obs}}. The \textit{scheduler} Python script reads from the PostgreSQL \textit{observation database}, which contains the details of survey and transient targets, and  then forms the current queue of pointings by calculating the highest priority pointing for each telescope in the network. It then communicates with the telescope pilots at each site through a simple Flask server, sending targets and receiving status information which feeds back into the scheduling algorithm. As observations are completed the scheduler updates the latest pointings in the queue, with the pilot regularly checking every 5 seconds for new priority targets.

New targets are added to the observation database by the \textit{sentinel} script from the \texttt{gtecs.alert} package \footnote{\url{www.github.com/GOTO-OBS/gtecs-alert}}. Event notices are received from NASA's Global Coordinates Network (GCN)\cite{GCN} through an Apache Kafka stream, with sources including the IGWN gravitational-wave detector network, the \textit{Fermi} and \textit{Swift} satellites, and the IceCube neutrino detector. The sentinel maps the localisation skymap from each notice onto the GOTO survey grid, to create targets to add to the observation database, where they will then be added to the queue by the scheduler. Other targets in the database include the regular all-sky survey, other more-focused secondary surveys and any user-added additions.

\section{HARDWARE CONTROL}
\label{sec:hardware}

The four GOTO telescopes are operated autonomously on-site by a \textit{pilot} master control program, part of the \texttt{gtecs.control} package \footnote{\url{www.github.com/GOTO-OBS/gtecs-control}}. The pilot issues commands to a series of \textit{hardware daemons} through the PyRO (Python Remote Objects) protocol, each daemon covering an aspect of the telescope hardware (cameras, mount, dome, etc). Throughout the night the pilot will run through a series of observing routines, sending commands to the daemons and receiving status updates. The pilot also monitors each daemon for any hardware errors, and has an inbuilt system of recovery commands to fix any common problems. The separate \textit{conditions} monitoring system ensures that the domes will close in bad weather, allowing the telescopes to operate without human monitoring.

As the pilots receive targets from the scheduler throughout the night, the resulting raw images are sent back to the GOTO servers at the University of Warwick, where they are processed by the Kadmilos pipeline \cite{2026Lyman}. Status information is also published by each pilot to a web dashboard and Slack channels, which can alert a human operator in case of any serious error that the pilot cannot fix itself.

\section{STATUS AND FUTURE PLANS}
\label{sec:future}
Since commissioning finished in 2023, the GOTO network has been fully operational as an autonomous network. It has successfully responded to hundreds of alerts, often observing within minutes of an event being detected.  To date GOTO has published over 5500 source discoveries to the IAU Transient Name Server \cite{TNS}.

With the system operating in a mature state, time can be invested in making the G-TeCS software packages maintainable and sustainable for long-term operations. Improving the package documentation and testing suites are both high priorities. New features, such as a QA feedback system and upgrades to the pilots, are also in development, as well as ongoing optimisations to the scheduling system.

Long term upgrades to the GOTO network include plans to replace the CCD detectors with new CMOS cameras, subject to future funding. Public data access is also a priority, with the ongoing development of the GOTO Lightcurve service\cite{Jarvis2026_lightcurve}.

\acknowledgments

The Gravitational-wave Optical Transient Observer (GOTO) project acknowledges the support of the Monash-Warwick Alliance; University of Warwick; Monash University; University of Sheffield; University of Leicester; Armagh Observatory \& Planetarium; the National Astronomical Research Institute of Thailand (NARIT); Instituto de Astrofísica de Canarias (IAC); University of Portsmouth; University of Turku; University of Birmingham; and the UK Science and Technology Facilities Council (STFC, grant numbers ST/T007184/1, ST/T003103/1 and ST/Z000165/1)

\bibliography{report}
\bibliographystyle{spiebib}


\end{document}